\documentclass[superscriptaddress,preprint]{revtex4-1}
\usepackage[utf8]{inputenc}
\usepackage{graphicx}
\usepackage[english]{babel}
\usepackage{amsmath}
\usepackage{geometry}
\usepackage{tikz}
\usepackage{url}
\usepackage{verbatim}

\usetikzlibrary{decorations.pathmorphing}
\usetikzlibrary{decorations.markings}
\usetikzlibrary{calc,intersections}
\usetikzlibrary{patterns}
\definecolor{purple}{rgb}{.5,0.,1}
\usepackage{pgf}
\usetikzlibrary{arrows}

\begin{document}

\title{Particle swarming of sensor correction filters}
 \author{Jonathan J. Carter}
\affiliation{Max-Planck-Institute for Gravitational Physics Callinstr. 38, 30167 Hannover, Germany}
 \affiliation{Institute for Gravitational Physics, Leibniz Universität Hannover, Callinstr. 38, 30167 Hannover, Germany}
  \affiliation{Institute for Gravitational Wave Astronomy, University of Birmingham, Birmingham, B15 2TT, United Kingdom}
 \author{Samuel J. Cooper}
 \affiliation{Institute for Gravitational Wave Astronomy, University of Birmingham, Birmingham, B15 2TT, United Kingdom}
 \author{Edward Thrift}
 \affiliation{Institute for Gravitational Wave Astronomy, University of Birmingham, Birmingham, B15 2TT, United Kingdom}
 \author{Joseph Briggs}
 \affiliation{SUPA Institute for Gravitational Research, Department of Physics and Astronomy, University of Glasgow, Glasgow G12 8QQ, UK}
 \author{Jim Warner}
 \affiliation{LIGO Hanford Observatory, Richland, Washington 99352, USA}
  \affiliation{LIGO Laboratory, California Institute of Technology, Pasadena, CA 91125, USA}
 \author{Michael P. Ross}
 \affiliation{Center for Experimental Nuclear Physics and Astrophysics, University of Washington, Seattle, Washington 98195, USA}
 \author{Conor M. Mow-Lowry}
 \affiliation{Institute for Gravitational Wave Astronomy, University of Birmingham, Birmingham, B15 2TT, United Kingdom}

\date{\today}

\begin{abstract}

Reducing the impact of seismic activity on the motion of suspended optics is essential for the
operation of ground-based gravitational wave detectors. During periods of increased seismic activity, low-frequency ground translation and tilt cause the Advanced LIGO observatories to lose `lock', reducing their duty cycles. This paper applies modern global-optimisation algorithms to aid in the design of the `sensor correction' filter, used in the control of the active platforms. It is shown that a particle swarm algorithm that minimises a cost-function approximating the differential RMS velocity between platforms can produce control filters that perform better across most frequencies in the control bandwidth than those currently installed. These tests were conducted using training data from the LIGO Hanford Observatory seismic instruments and simulations of the HAM-ISI (Horizontal Access Module Internal Seismic Isolation) platforms. These results show that new methods of producing control filters are ready for use at LIGO. The filters were implemented at LIGO's Hanford Observatory, and use the resulting data to refine the cost function.

\end{abstract} 
\maketitle

\section{Introduction}
The first observation of gravitational waves from a binary black hole \cite{PhysRevLett.116.061102} was just a few years ago, but there has since been 11 confirmed astrophysical signals during the O1 and O2 observing periods \cite{PhysRevX.9.031040}, and many more event candidates during O3 \cite{chirp}. For the Advanced LIGO detectors \cite{aligo,PhysRevLett.116.131103}, the increased frequency of detections has placed a greater emphasis on the stability and consistency of the observatories.

The extraordinary sensitivity of the detectors places stringent requirements on the residual vibration in the measurement band \cite{Robertson_2002}. Additionally, for the observatories to enter into an operational state, the mirrors must be `locked' to an ideal operating point  \cite{martynov2015lock,datrier19Wind,RossLockLogbook}. Low frequency ground motion can cause the mirrors to move out of this operating point, a process that is known as loosing lock, if this motion is not sufficiently suppressed. To reduce low-frequency vibrations, a complex multi-stage active isolation system is employed \cite{isibsc2,isibsc}. The active control employs an array of differential and inertial sensors on each stage. At frequencies between 0.3 and 15\,Hz, feedback control provides most of the vibration suppression; near the microseismic peak (50-300\,mHz)\cite{doi:10.1098/rsta.1950.0012} it is done by feed-forward control.

`Sensor correction' (SC) is a control technique where signals from ground seismometers are filtered and summed with the output of differential displacement sensors that measure the position of LIGO's isolated platforms, as shown in Figure~\ref{fig:scPath}. This removes the ground contribution of the displacement sensor to yield the inertial platform motion.  The aim is to use the excellent noise performance of vault-grade seismometers at low frequencies. However, there is a cut-off below 0.1\,Hz where ground seismometers are dominated by their tilt susceptibility \cite{Lantz09}.  This noise is avoided by using the displacement sensors as the feedback signal. A good sensor correction filter aggressively tackles the translational ground motion without including frequencies where noise dominates, in particular, tilting contamination. 

This paper presents a novel approach for improving sensor-correction filter design \cite{derosa2012global,hua2005low} based on particle swarm optimisation. Improved filters are found with an unguided search using real data, for a variety of environmental conditions. By using real data-streams  from all three types of installed sensors, and applying IIR filters exactly as LIGO's real-time control system does, the tool becomes immune to a slew of systematic issues including finite coherence, differing transfer-functions, and numerical implementation effects. A cost function was designed to mimic the physical quantity we believe is most important for LIGO's performance, the RMS inertial velocity. The velocity spectrum was shaped in an a-priori fashion to account for known physical effects of common-mode rejection and resonances of the suspended optics. Finally, the search variables were reparameterised to remove degeneracy in the search space.

In section 2, comparisons between current filter designs methods and this technique are discussed. Section 3 details the implementation of the particle swarm. Section 4 reviews the filters and their performance when installed at the Hanford site.

\section{Numerically optimised filter design}
Filter design is a challenge in many fields; each of which have their own relevant filter design methods. The SC filters used at LIGO are hand-tuned by the operators and commissioners. While this can be an effective design method, it requires substantial experience and an understanding of the performance requirements. Until now there has not been a quantitative performance metric, and it has been difficult to evaluate the relative quality of different filters. Furthermore, there exists the possibility that the designs are limited by human biases. Additionally, since this design and testing process is time consuming, the filters are only replaced if there are  obvious performance issues.
\par
Other fields have developed a wide array of tools  for generating of control filters to suit their specific needs.  An example often used for motion suppression in car suspensions is reinforced learning automata, such as described by Howell \textit{et al.} \cite{HOWELL1997263} and other later groups\cite{Kashki2010383}. Practical limitations prevent this method from direct application to gravitational wave detectors. It would require in operation testing so that the direct output of the filter can be evaluated. In operation measurements are not feasible for a gravitational-wave detector, due to the unacceptable loss of observation time of the testing process; this strongly motivates the development of rigorous off-line testing procedures.
\par
Storn \textit{et al.} \cite{storn1996} pioneered a method for when there is \textit{a priori} knowledge of the desired filter response. They use a differential evolution optimiser to fit poles and zeros to a prescribed phase and magnitude response, whilst employing cost `penalties' in order to meet stringent stability conditions. 
This method is not well suited for designing a sensor-correction filter as the desired response is unknown due to the complex noise dynamics and limited coherence of the sensors in the isolation system.

Other groups have used particle swarms to design control filters based upon Storn's work \cite{Krusienski2004,Luitel2008}. They show that the particle swarm optimisation algorithm can effectively design filters due to its ability to sample large parameter spaces. A direct comparison between genetic algorithms and particle swarms is presented in \cite{Krusienski2004}. The particle swarm reached a lower minimum more consistently, though there was little difference between the two methods; a similar conclusion was reached during the development of the optimisation method presented here. 
Modifications to the particle-swarm optimiser are described in \cite{Luitel2008} that made it similar to the ``branch and prune" optimisation technique. The changes significantly improved the optimisation performance, and this illustrates the relative flexibility and adaptability of the particle-swarm technique.
\par
A numerical optimiser enables a conceptually simple way to design SC filters.
The only bound on the solver are the asymptotes outside the band of interest (4\,mHz-4\,Hz), which are entirely controlled by the number of poles and zeros in the filter. These asymptotes have strong physical motivation, however. Within the entire frequency band of interest a completely unguided and randomly seeded search is made. This allows for non-obvious, but beneficial features to be incorporated in the filter. 
Although conceptually simple, implementing such an optimiser proved to be a non trivial task.
\par
Data collected from the wide array of on-site sensors was used for the calculation of filter performance during generation. 
Calculating the overall ground injection using this numerical, time domain approach allows for the inclusion of finite coherence between multiple sensors preventing the over-estimation of the sensor correction filter. Furthermore, many real world effects of implementing filters such as differing transfer functions between sensors are already incorporated in the resultant filters. Because the data processed is in the time domain, only real coherence can be subtracted. 
Finally, doing so with time domain data is the numerically closest operation to on site action. This means an unstable filter will be inherently penalised as it will cause a large increase motion in the system.
\par
While this work focuses on sensor correction filters, the tools modular design means that it can be adapted to design various control filters within the seismic isolation system. 
The problem for filter generation then becomes purely a matter of finding an appropriate cost function.
 
\section{Guiding the swarm}
All numerical optimisation processes require a cost function. Using a particle swarm optimisation routine allows for a computationally intensive one. The cost function mimics the implementation of the sensor-correction filter performed by the LIGO's real-time control system. From there, the filter's performance was analysed to estimate the disturbances injected into LIGO's isolated platforms. Additional weighting was applied to account for key features such as suspension resonances. 

\subsection{Generating sensor-corrected signals off-line} \label{sec:cost}

\begin{table}
\centering
 \begin{tabular}{||c  c  c||} 
 \hline
 Sensor Name & Measures & Contributions  \\ [0.5ex] 
 \hline\hline
  BRS & Ground inertial rotation & $\tilde{\theta}_{\rm{g}}+\tilde{n}_{\rm BRS}$\\  \hline
  STS2 & Ground inertial translation & $\tilde{\dot{x}}_{\rm{g}}+\tilde{n}_{\rm STS2}+\tilde{\theta}_{\rm{g}}\frac{g}{\omega}$\\ \hline
  CPS & Relative platform velocity & $\tilde{\dot{x}}_{\rm{p}}-\tilde{\dot{x}}_{\rm{g}}+\tilde{n}_{\rm CPS}$\\ \hline
  T240 & BSC-platform inertial translation & $\tilde{\dot{x}}_{\rm{p}}+\tilde{n}_{\rm T240}+\tilde{\theta}_{\rm{p}}\frac{g}{\omega}$\\ \hline
  GS13 & HAM-platform inertial translation & $\tilde{\dot{x}}_{\rm{p}}+\tilde{n}_{\rm GS13}+\tilde{\theta}_{\rm{p}}\frac{g}{\omega}$\\ \hline
 
 \end{tabular}
\caption{The instruments used in the construction of the sensor correction cost. The names of the sensors are: BRS - Beam Rotation Sensors, STS2 - Streckeisen STS-2 force-feedback seismometer, CPS - Capacitive Position Sensor, T240 - Trillium T240 force-feedback seismometer, and GS13 - Geotech Instruments GS-13 short-period seismometer. }
\label{tab:references}
\end{table}
\begin{figure}
	\centering
	\includegraphics[width=0.9\linewidth]{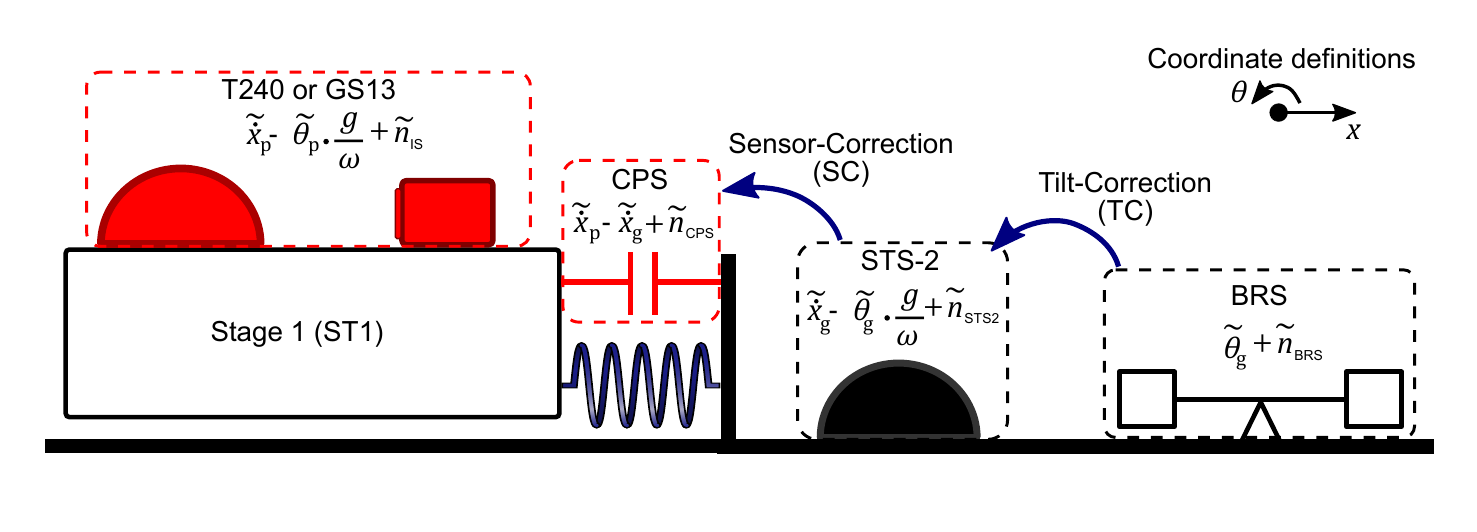}
	\caption{The signal path of tilt- and sensor-correction. Tilt correction is used at LIGO Livingston (LLO) in the beam-direction for all chambers, while it is only used at LIGO Hanford (LHO) for the end stations (ETMX, ETMY). Sensor correction is used on all chambers at both observatories in translational degrees of freedom (X,Y,Z). The roles of each sensor are described in Table~\ref{tab:references}. Figure adapted from Cooper \cite{Cooper19PhD} with permission.}
	\label{fig:scPath}
\end{figure}

Figure~\ref{fig:scPath} shows a functional schematic of the sensor correction signal path. Each block represents a type of sensor installed in the LIGO observatories and shows the contributions to their output signals in terms of inertial translation, tilt (or rotation), and self-noise. All variables carry a tilde to indicate the Fourier transform. From this block diagram, the sensor-corrected CPS signal, $\tilde{a}_{\rm{sc}}$, can be calculated based on the signals from the seismometer and rotation sensor on the ground

\begin{eqnarray}
\tilde{\theta}_{\rm r} & = & \tilde{\theta}_{\rm g} - {\rm TC}(\tilde{\theta}_{\rm g} + \tilde{n}_{\rm BRS}) \\
\tilde{a}_{\rm{sc}} & = & \tilde{\dot{x}}_{\rm{p}} - \tilde{\dot{x}}_{\rm{g}} + \tilde{n}_{\rm CPS} + {{\rm{SC}}}(\tilde{\dot{x}}_{\rm{g}} - \tilde{\theta}_{\rm r} \frac{g}{\omega}), \label{eqn:totalInjection} \\
\tilde{a}_{\rm{sc}} & = & \underbrace{\tilde{\dot{x}}_{\rm{p}}}_{\rm{Platform\ Motion}} + ~ \tilde{n}_{\rm CPS} - \underbrace{(1-{{\rm{SC}}})\tilde{\dot{x}}_{\rm{g}}}_{\rm{Ground\ Injection}} - ~ \underbrace{{{\rm{SC}}}(\tilde{\theta}_{\rm r}\frac{g}{\omega}  + \tilde{n}_{\rm STS2})}_{\rm{Tilt\  Injection}}. \label{eqn:sensorCorrection} 
\end{eqnarray}
Here $\tilde{\dot{x}}$ denotes translational motion, $\tilde{\theta}$ tilt motion, while subscripts $\rm{g}$ and $\rm{p}$ show whether this is associated with ground or platform motion respectively. The complex frequency responses of the sensor-correction and tilt-correction filters are denoted by ${\rm SC}$ and ${\rm TC}$ respectively, $\omega$ is the angular frequency, and $g$ is acceleration due to gravity. The `residual tilt', $\tilde{\theta}_{\rm r}$, is equal to the ground tilt in degrees of freedom where there is no BRS, as described below. The $\tilde{n}$ terms denote the self-noise for the instrument indicated in the subscript. Table~\ref{tab:references} describes the role of each instrument and shows the contributions that make up the output signals. 


The ground and platform inertial motions are measured by seismometers. The ground rotation can only be measured where there is a Beam Rotation Sensor (BRS) \cite{venkateswara2017subtracting}. At the time of writing, there were four BRS instruments at LIGO Livingston observatory, measuring the rotation that affects degrees of freedom parallel to the beam-line direction at the corner-station and at both end stations. At LIGO Hanford observatory there were two BRS instruments, one in the beam-direction at each end station. 

In the ideal case, the sensor-corrected CPS signal only includes the platform motion, which can subsequently be suppressed by feedback control. By constructing the sensor-corrected CPS signal offline from long sections of data, many different sensor-correction filters can be implemented and tested without requiring valuable operating time of the LIGO observatories.
Signals are processed to flatten the frequency response in band and are converted to velocity and radians for translation and rotational motion respectively.

The optimisation problem of the sensor-correction filter is now clear: we must add as much of the inertial ground-motion term as possible to the CPS signal, while minimising the injection of tilt and inertial-sensor self-noise. The compromise between the two terms in Equation~\ref{eqn:sensorCorrection} is complicated further by the small frequency separation of important dynamics in their spectra. The dominant contribution to the RMS ground translation occurs between 0.1 and 0.3\,Hz (the secondary micro-seismic peak), and the tilt-injection begins to dominate the seismometer signals below approximately 0.05\,Hz \cite{windLLO,doi:10.1063/1.4862816,windproofingLIGO}. 

{The sensor correction filter can be thought of as one part of a complementary pair of `blending' filters \cite{kissel2010PHD}, an example of sensor correction filters can be found in figure \ref{fig:csSCFilters}. The sensor-correction filter attenuates tilt and inertial sensor noise, while the ground injection is reduced by a factor of one minus the sensor-correction filter. When installed into the detector only the sensor correction filter is installed, the filter complement is used to show the ground injection suppression. 
}

The optimisation is restricted to a band between 4\,mHz and 4\,Hz. Below 4\,mHz, the effect of all inertial sensors is negligible on the platform RMS velocity \cite{windproofingLIGO}. Above 4\,Hz, the CPS contribution to the platform feedback signal becomes negligible~\cite{isibsc2,isibsc}, due to the suppression provided by the high and low pass blending filters respectively.

\subsection{A physically motivated cost function}




The design of a cost function often encodes much of the complexity of an optimisation problem. To aid this process, we created a cost that was the integral over frequency of the tilt-injection and ground-injection terms identified in Equation~\ref{eqn:sensorCorrection}. The ability to directly compare these spectral integrands with spectra of the input signals was of great practical benefit during debugging, and helped to shape both the cost function and performance penalties described below. The final cost is also approximately equal to the residual RMS velocity of the platform.


The RMS velocity of the platform is a useful figure of merit because it is correlated with the scattered light performance of the interferometer. It also balances the need to control drift at low frequencies (better evaluated by the RMS position) and limiting control forces at high-frequencies (better evaluated by RMS acceleration). 

\subsection{Tilt injection}

\begin{figure}
	\centering
	\includegraphics[width=0.9\linewidth]{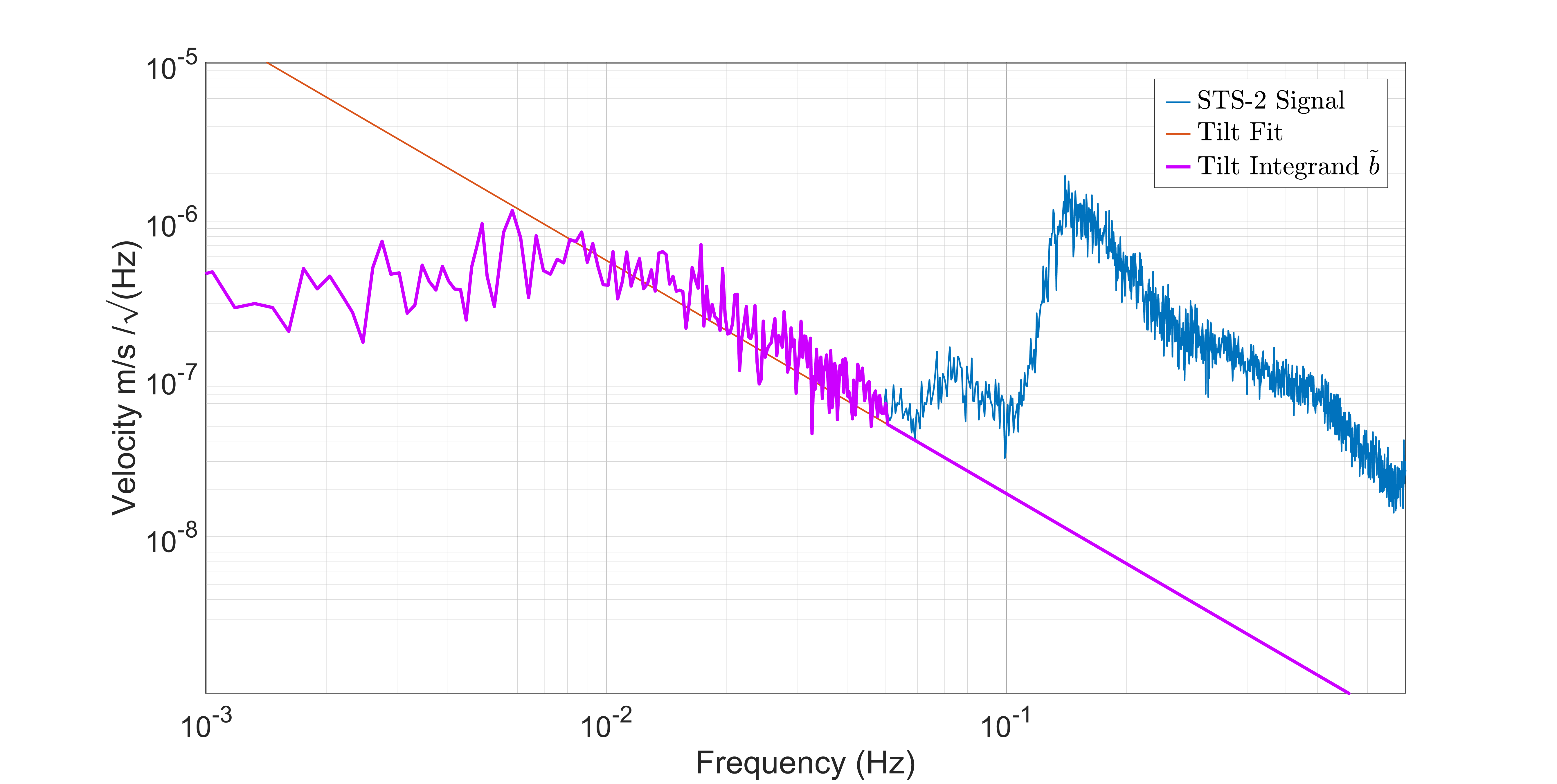}
	\caption{Figure showing the estimation of the tilt integrand, $\tilde{b}$, used in the swarming process. Between 0.01\,Hz and 0.05\,Hz the STS-2 motion is used, as this is strongly correlated with the local wind speed. A linear fit is used in this region to estimate the tilt coupling for the rest of the frequency band. }
	\label{fig:stsTiltFit}
\end{figure}

At frequencies below approximately 50\,mHz, the ground seismometer signal becomes completely dominated by tilt caused by wind pressure acting on the buildings \cite{windproofingLIGO}. Furthermore, the STS2 is AC-coupled at 8\,mHz, and its output is no longer an inertial measure of the ground velocity. Therefore the STS2 signal below 50\,mHz is considered to be part of the tilt-injection spectrum. Above this frequency, a simple power-law extrapolation of the tilt spectrum is made, as shown in figure~\ref{fig:stsTiltFit}. The small region from 50-100\,mHz, which contains the primary micro-seismic peak, contains little spectral power and is de-weighted by the ground weighting function shown in figure~\ref{fig:swarmCostWeight}.

The contribution to the total cost function is therefore the integral of the tilt-injection spectrum multiplied by the frequency response of the sensor-correction filter.


\subsection{Ground injection}

With no sensor-correction filter applied, the inertial ground motion can be completely transferred to the platform through the CPS sensor. It is therefore important to reduce this term as much as possible, especially at the secondary micro-seismic peak, seen between 0.1 and 0.3\,Hz in the plots below. 

\begin{figure}
	\centering
	\includegraphics[width=0.9\linewidth]{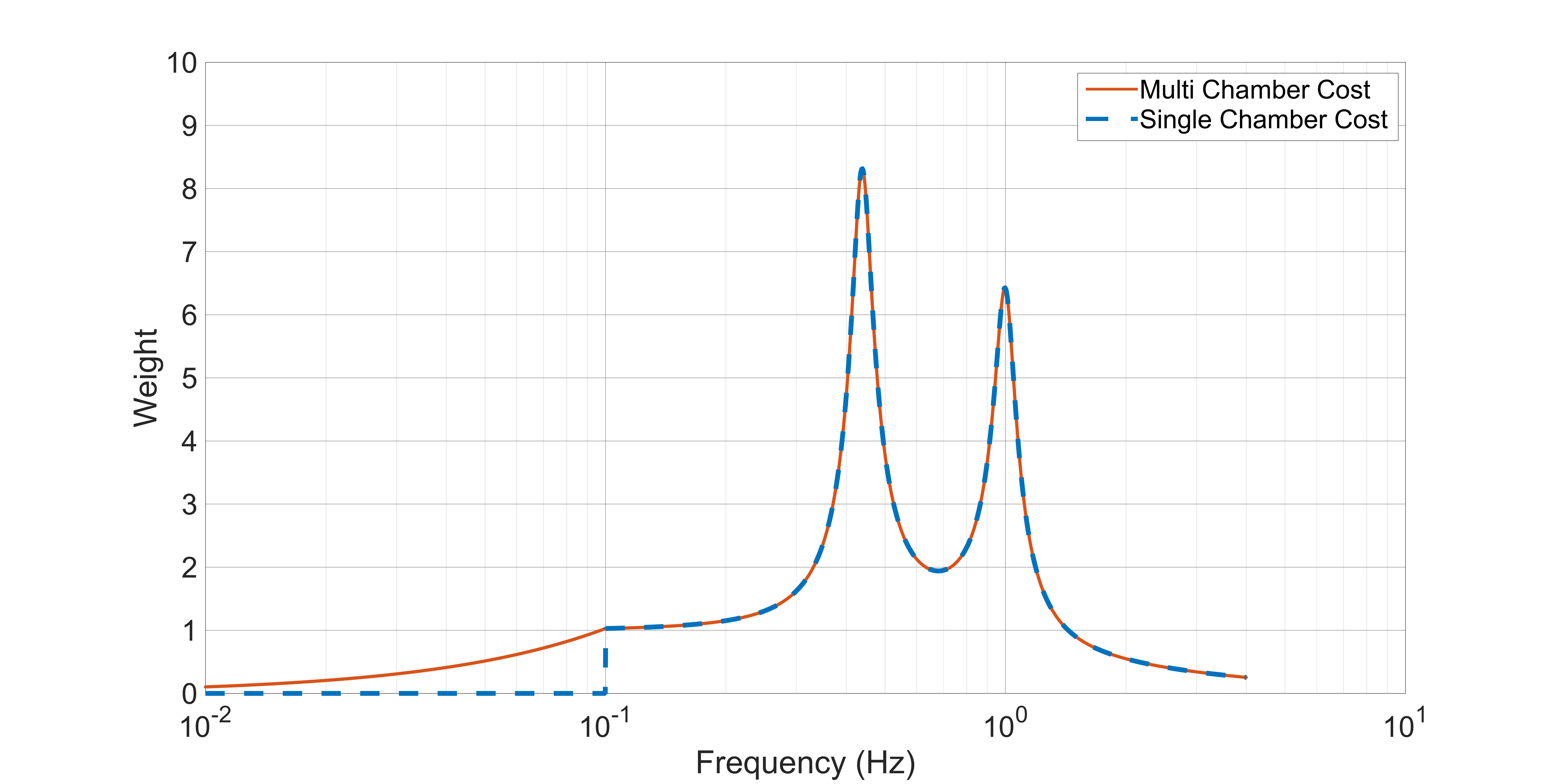}
	\caption{Weighting function, $W_{\rm{g}}$, for the ground-injection spectral contribution to the cost function. Peaks at 0.45\,Hz and 1\,Hz correpsond to suspension resoances.}
	\label{fig:swarmCostWeight}
\end{figure}

Each isolation platform has high Q, multi-stage suspensions that further reduce vibration above approximately 1\,Hz. The lowest frequency pendulum-modes of the `quadruple' suspensions, at 0.45\,Hz and 1\,Hz, are responsible for a significant fraction of the residual motion in LIGO, and it was important to include these dynamics in our cost. The weighting shown in Figure~\ref{fig:swarmCostWeight} includes a simplified and broadened fit of the suspension response.

At frequencies below 0.1\,Hz the ground injection cost is linearly de-weighted to zero. There are two reasons for this: first, the spectrum of the ground seismometer is increasingly dominated by tilt, and second, actual inertial translation is almost completely common-mode, even down the 4\,km long arms of LIGO. A vanishingly small actuation force is needed to combat the differential component. At frequencies above 1\,Hz, the CPS sensor plays very little role in the motion of the platform, and the cost is de-weighted to improve the numerical precision of the optimisation in the critical 0.1-1\,Hz region.

There are therefore two components to the final cost of the swarming process, the ground and tilt injection. The ground injection signal is the previously calculated sensor corrected CPS signal minus the inertial platform motion, $\tilde{\dot{x}}_{\rm{inj}}$. This term is calculated in the time domain to account for the finite coherences between the ground seismometer, CPS, and inertial sensor on the platform. The tilt injection term is calculated in the frequency domain due to the estimation processed used to calculate the tilt integrand, $\tilde{b}$ above 50\,mHz. This is multiplied by the frequency response of the sensor correction filter producing the tilt injection term.

The ground injection term, $\tilde{\dot{x}}_{\rm{inj}}$, is given by Equation~\ref{eqn:groundInjection} and is defined here in the frequency domain. $\tilde{a}_{\rm{sc}}$ is the previously calculated sensor corrected CPS, $\tilde{\dot{x}}_{\rm{p}}$ is the platform motion, $\tilde{n}_{\rm{IS}}$ is the noise of the inertial sensor, this is either a GS13 or a T240 and $\tilde{\theta}_{\rm{p}}$ is the platform tilt. In reality, the swarming process calculates this in the time domain to ensure the swarmed filter is stable. The tilt injection term, $\tilde{\theta}_{\rm{inj}}$, is given by,
\begin{eqnarray}
\tilde{\dot{x}}_{\rm{inj}} & = & \widetilde{a}_{\rm{sc}} - \dot{\tilde{x}}_{\rm{p}} + \tilde{n}_{\rm{IS}}+\tilde{\theta}_{\rm{p}}\frac{g}{\omega} \label{eqn:groundInjection},\\
\tilde{\theta}_{\rm{inj}} & = & {\rm{SC}} ~ \tilde{b}, \label{eqn:tiltinjection}
\end{eqnarray}
and is calculated in the frequency domain. Here the tilt fit curve, $\tilde{b}$, shown in Figure~\ref{fig:stsTiltFit}, is multiplied by the sensor correction filter and the ground injection term is multiplied by the ground weighting term, $\rm{W_g}$, shown in Figure~\ref{fig:swarmCostWeight}. The tilt and ground injection terms are both integrated over the entire frequency band and summed in quadrature producing the final cost, given by

\begin{eqnarray}
\rm{cost}^2 & = & \int_{0.004\,\rm{Hz}}^{4\,\rm{Hz}} |\rm{W_{g}} ~ \tilde{x}_{\rm{inj}}|^2 ~df + \int_{0.004\,\rm{Hz}}^{4\,\rm{Hz}} |\tilde{\theta}_{\rm{inj}}|^2 ~df. \label{eqn:finalCost}
\end{eqnarray}
Using on-site data provides many advantages towards filter design. 
With the sensor correction filter, the output of the platform seismometer acts as an in-situ test of the quality of the filter. The goal of sensor correction is to remove the ground motion in the CPS, such that the CPS follows the platform motion creating a quasi-inertial sensor. The T240 measures this and can therefore be used as a measure of quality, when factoring in the noise and tilt as measured by this sensor.

The designed filter can then be evaluated using other sets of input motion to verify their performance in a range of environmental conditions. Ensuring the cost function outlined above is physically motivated allows for quick comparisons between the performance of multiple filters and filters across different sets of environmental conditions.

\subsection{Particle Swarm Optimisation}
\label{sec:particleSwarm}

Particle swarm routines aim to simulate social behavior to explore a parameter space. 
A number of `particles' are generated with randomised parameters.
In the context of this work, these contain the information necessary to build a control filter Zero Pole Gain (ZPK) function.
The routine then calculates a cost for each of the particles using the cost function. 
In this case it must also build the ZPK of the filter from the generated parameters and then calculate the cost using the relevant equations discussed in section \ref{sec:cost}. 
Particles then determine how to change their parameters based on the overall global minimum, the local minimum, and a random `kick'.
The process is then repeated until an exit condition of the swarm is met.
During initial testing a swarm size of 500-1500 particles was able to sufficiently explore the parameter space whilst still enabling many iterations to be performed in a reasonable time frame on a typical laptop. 
\par

For use in LIGO, each designed sensor correction filter should use no more than 20 second order sections, this means the swarmed sensor correction filter can only contain 4 complex poles and 3 complex zeros. The filters are seeded with a single real pole and three real zeros at 0\,Hz to shape the filter and guarantee the required roll off at low frequency. 

The use of a particle swarm allows for an unguided search of the created parameter space. It is forced to construct the initial generation of filters with random parameters for the frequency and Q (for complex poles and zeros) so that the parameter space for any potential filters is fully explored. 

\subsection{Reparameterisation}

How the parameter space is explored by the optimisation routine was the first area considered when refining the filter generation tool.
By choosing the space in which the roots are generated, the weighting of filter construction in early generations can be skewed to allow for faster and more reliable convergence.
The positions of roots in a ``ideal" filter will typically not be distributed uniformly. 
If the filters were generated in a linear frequency space, where the value generated is the position of the root, then shaping the low frequency response in a wide data range would take a much greater optimisation time. 
Instead, to account for various physical factors in certain frequency bands, roots will typically be spread over a logarithmic scale with a high density at crossover points to allow for better shaping. 
\par
The degeneracy of the space is the largest problem with a natural frequency creation, (where the value of the particle is the frequency of the root).
Any two roots of a polynomial can be interchanged and the resulting transfer function will remain the same, displayed by Figure~\ref{fig:noedge} (top). 
This artificially inflates the parameter space by a factor proportional to the number of roots factorial.
Larger parameter spaces will take longer for the optimisation routine to fully explore.
To solve this problem, roots were ordered as shown in Figure~\ref{fig:noedge} bottom.
Here the lowest frequency root is defined in natural frequency space and the subsequent roots defined in terms of the difference to the last one generated.
This orders the roots, solving the degeneracy problem.
Ordering of the roots presents different problems to the filter construction in the form of imposing boundaries on root placement.
To maintain the appropriate slopes outside of the studied region, no poles or zeros can be placed there.
When building a filter, the range of acceptable poles and zeros should therefore be bound to the range of frequencies that the data covers. 
However, by defining the relations between roots, no boundary conditions can be imposed as shown by Figure~\ref{fig:noedge}, where each jump is less than the range of frequencies allowed, but still the final root lies outside the acceptable range.
This wastes a large amount of the available computing power on filters which must be discarded.
\par
With effort on two fronts this problem was overcome. 
The first thing was to add a large scaling cost when an invalid filter was produced. 
This extra cost was based on a multiplicative term of how many roots and how far each root was outside the acceptable range.
 This created a ``bucket" in the parameter space which allowed particles to drift back into the acceptable range by imposing a gradient towards acceptability.
The second tactic was a careful initialisation of the swarm matrix. 
Before the swarm was run, a first generation of filters was created in natural frequency space and then ordered from highest to lowest. The difference between these was then calculated and used as the parameters for generation in the swarm. This meant initial generation of particles were all viable filters and any particle which drifted out of range was pushed back by the aforementioned bucket. A graph showing the convergence of the different methods discussed is shown in Figure~\ref{fig:convtest}, showing that after 120 generations with 500 particles the ordered generation initialised jumps achieved a significant improvement in best cost. It showed that both steps were necessary to see an improvement.
\par 
The parameters used by the swarm to make a control filter are shown in Table~\ref{tab:params}. 
\begin{figure}
    \centering
    \begin{tikzpicture}[>=latex,scale=1.5,decoration=snake]

        \draw[thick](1,4)--(1,3)--(0,3)--(10,3)--(9,3)--(9,4);
        \draw[thick](1,4)--(1,3)--(0,3)--(10,3)--(9,3)--(9,4);
        \fill[black](2,3.5)circle(0.1);
        \fill[black](4,3.5)circle(0.1);
        \fill[black](6,3.5)circle(0.1);
        \fill[black](8,3.5)circle(0.1);
        
        \draw[thick](2,3.5)node[above]{$f_1$};
        \draw[thick](4,3.5)node[above]{$f_2$};    
        \draw[thick](6,3.5)node[above]{$f_3$};
        \draw[thick](8,3.5)node[above]{$f_4$};
        
        \draw[thick](5,3)node[below]{Frequency Space};
        \draw[thick](1,3)node[below]{$f_{\rm{{min}}}$};
        \draw[thick](9,3)node[below]{$f_{\rm{max}}$};
        
        \draw[ triangle 45 -triangle 45 , thick](2.1,3.5)--(3.9,3.5);
        \draw[thick](3,4)node[above]{Two nodes are interchangeable};  
        
        \draw[thick](1,1)--(1,0)--(0,0)--(10,0)--(9,0)--(9,1);
        \draw[thick](1,1)--(1,0)--(0,0)--(10,0)--(9,0)--(9,1);
        \fill[black](2,0.5)circle(0.1);
        \fill[black](5,0.5)circle(0.1);
        \fill[black](8,0.5)circle(0.1);
        \fill[black](0,0.5)circle(0.1);

        \draw[thick](8,0.5)node[right]{$f_1$};
        \draw[thick](3.5,1.5)node[above]{$\Delta f_1$};
        \draw[thick](6.5,1.5)node[above]{$\Delta f_2$};
        \draw[thick](1,1.5)node[above]{$\Delta f_3$};
        
        \draw[thick](5,0)node[below]{Frequency Space};
        \draw[thick](1,0)node[below]{$f_{\rm{min}}$};
        \draw[thick](9,0)node[below]{$f_{\rm{max}}$};
        
        \draw[-triangle 45, thick, dashed](8,.5)arc(35:144:1.8);
        \draw[-triangle 45, thick, dashed](5,.5)arc(35:144:1.8);
        \draw[-triangle 45, thick, dashed](2,.5)arc(35:144:1.2 and 1.8);

    \end{tikzpicture}

    \caption{(top) The problem of roots degeneracy where $f_1$ and $f_2$ can be switched and the resultant filter remain the same. (bottom) How the roots of a polynomial can escape the frequency range of desired generation with jumps of size less than the frequency range when ordered generation is used.}
    \label{fig:noedge}
\end{figure}
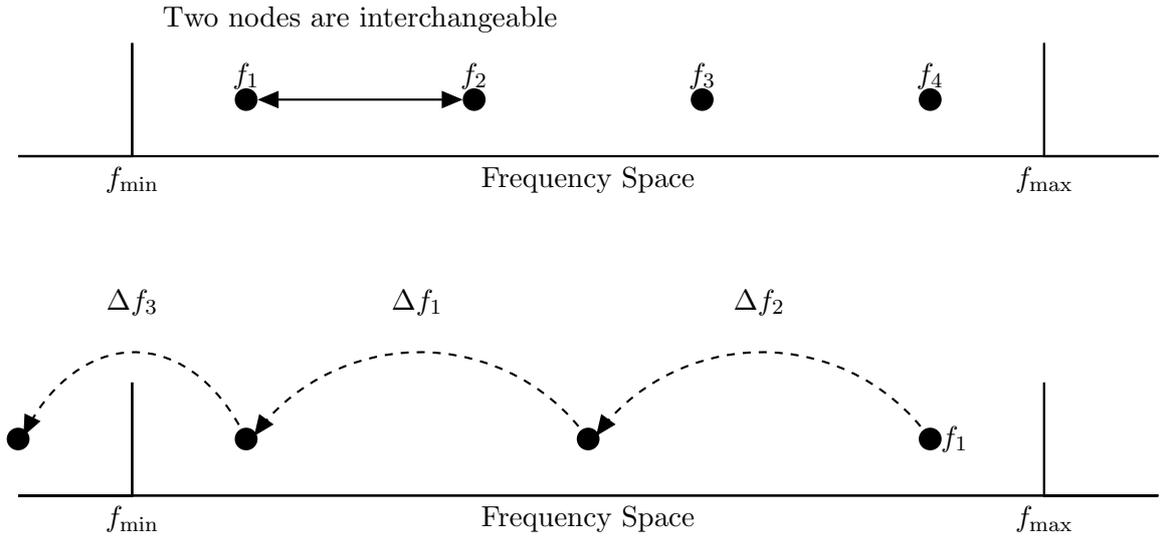
\begin{figure}
    \centering
    \includegraphics[width=0.9\linewidth]{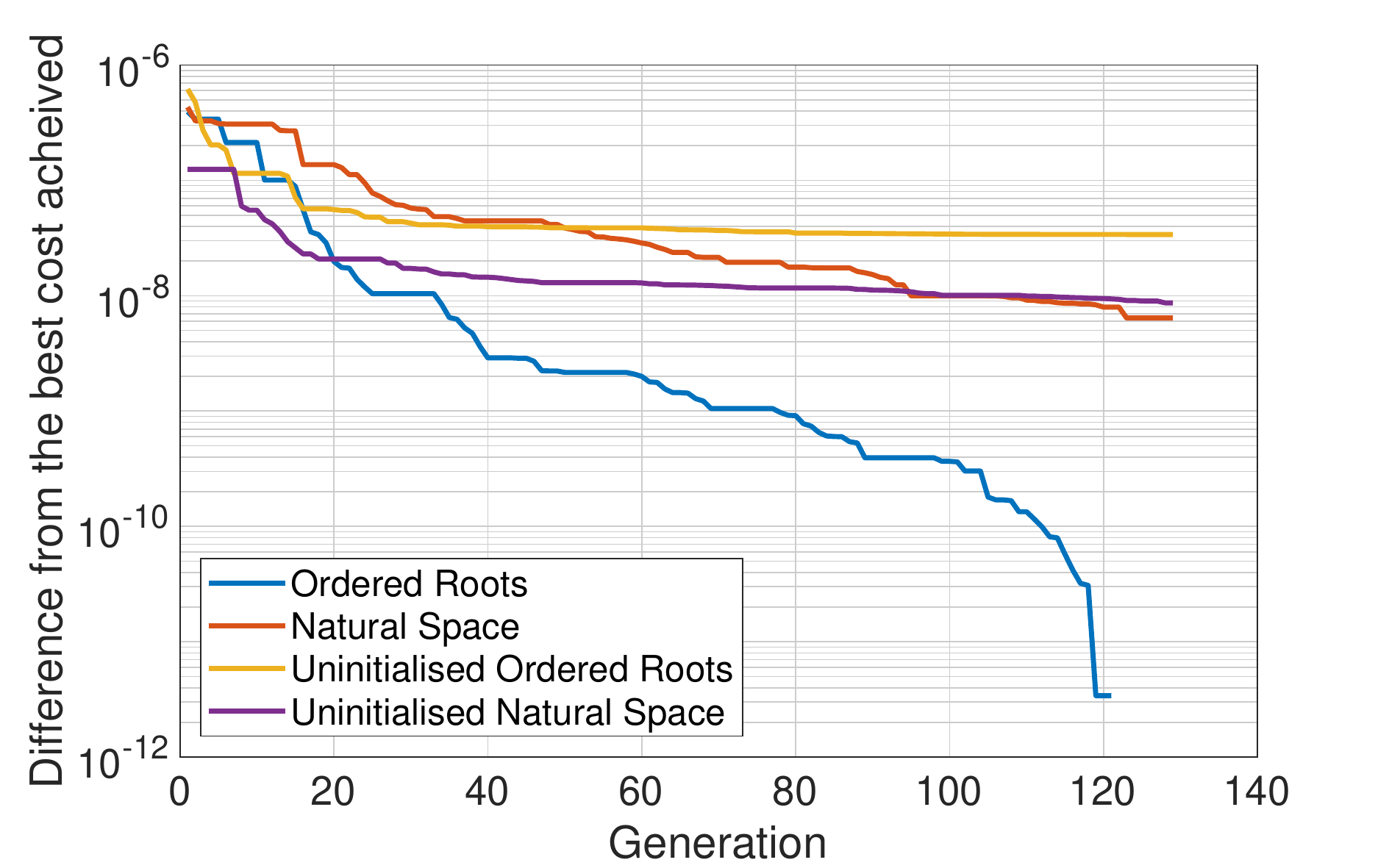}
    \caption{The convergence to best filter of the different parameter spaces (ordered and natural) and the effect of swarm initialisation. All data is measured relative to the best filter. This corresponds to a 13$\%$ improvement in frequency space. The tests were run several times and those with similar cost in initial generation displayed. }
    \label{fig:convtest}
\end{figure}

\begin{table}
\centering
 \begin{tabular}{||c c c c||} 
 \hline
 Parameter & Description & Lower Limit & Upper Limit \\ [0.5ex] 
 \hline\hline
  1 &  Gain & 0.95 & 1.05 \\ \hline
  2 & $\log_{10}(f)$ of real pole & $\log_{10}(f_{\rm{min}})$ & $\log_{10}(f_{\rm{max}})$ \\ \hline
  3 &  $\log_{10}(f_1)$ of highest complex pole & $\log_{10}(f_{\rm{min}})$ & $\log_{10}(f_{\rm{max}})$ \\ \hline
    4-7 &  $\log_{10}(\Delta f_{1-3})$ from last complex pole & $ \quad \log_{10}(f_{\rm{min}}/f_{\rm{max}})$ & 0 \\ \hline
   8 &  $\log_{10}(f_1)$ of highest complex zero & $\log_{10}(f_{\rm{min}})$ & $\log_{10}(f_{\rm{max}})$\\ \hline
 9-11 &  $\log_{10}(\Delta f_{1-3})$ from last complex zero & $ \quad \log_{10}(f_{\rm{min}}/f_{\rm{max}})$ & 0 \\ \hline
    12-20 & Quality factors for complex roots & 0.35 &5 \\ \hline
 \end{tabular}
\caption{A list of the parameters sent to the swarm and how they are converted into a filter function. Each frequency and $Q$ is used to create one complementary pair of complex roots. The range of $Q$ values allow the poles and zeroes to become slightly over-damped, making a pair of real poles. }
\label{tab:params}
\end{table}

\section{Implementation at LIGO Hanford Observatory}
\subsection{Choice of training data}
The data used for filter generation was collected during the commissioning break between the second and third observing runs, when the detector was in a damped only state (the isolation forward control loops were not in in operation). This simplifies the sample data as the many control techniques did not have to be considered in the processing of the input data, nor was it necessary to account for closed loop stability requirements. 

\par

The swarmed filters were created using data collected during times of above average environmental disturbance, it is assumed that if the filter was designed during these times it would be more robust in a wider range of environmental conditions. 
Since the detectors have the ability to switch between different filters, it would be possible to generate filters to account for seasonal variations of ground motion and wind speed, though this is outside the scope of the paper.

\subsection{Optimising Multiple Chambers}

In the corner stations of both LIGO Observatories one sensor-correction filter is applied to the STS-2 signal in each degree of freedom, and distributed to all chambers. This ensures that the residual tilt injection that couples through the sensor correction path is the same for each of the isolation platforms, reducing differential platform translational motion~\cite{Cooper19PhD}(Chapter 6). Therefore when optimising the sensor-correction filters for the corner station, a filter must be produced that is effective in all chambers. To produce the required input data for the swarming process, signals from each chamber on the `beam axis' were averaged together. For each of the horizontal translational degrees of freedom (X and Y) we combine the signals from all the relevant chambers in the main building.

\subsection{Comparison with current filter}


\par
The filters produced by the swarm were tested at the Hanford site between the 11th of October 2019 and 29th October 2019. The observatory had several other events occurring that resulted in the ISI (Internal Seismic Isolation) platforms operating at atmospheric pressure leading to overall higher noise. On site operators noted that the filter injected more ground tilt into the ISI, this is due to the filter having a factor 6 higher gain at low frequency compared to the previous sensor correction filter. 
\par

The direct comparison of two sensor correction filters that were active during two different stretches of time is complex due to the differences in the input ground and wind conditions. As a result, the cost function that was used to design the filters would not be appropriate. Moreover, the GS-13, used on the HAM ISIs as a witness sensor is limited by its own self noise below approximately 0.05\,Hz, is susceptible to platform tilt, and thus would not be a good witness sensor below this frequencies.
\par
As such, below approximately 0.1\,Hz the performance of the filters was determined by the sensor corrected CPS signal and by the GS-13 or T-240 above this frequency. The exact form of the figure of merit is given by Equation~\ref{eqn:FOM}. The sensor corrected CPS provides information on the relative tilt injection and partially the suppression of the microseismic peak while the on-platform inertial sensor measures the suppression of the microseismic peak and the isolation performance above 0.1\,Hz. 


\par

To quantify the performance of each sensor correction filter, the amplitude spectral density was taken of each sensors in table \ref{tab:references}, whereby the signals are converted into displacement and are plant inverted where appropriate. The signal from the CPS from 1\,mHz is stitched with the on-platform inertial sensor, IS, either the GS13 or T240 as appropriate, at 0.1\,Hz as shown by, 
\begin{eqnarray}
\tilde{\dot{x}}_{m} =
\begin{cases}
	 {\rm CPS}, & 0.001 \leq f \leq 0.1 \\
	 {\rm IS},  & 0.1 < f \leq 1.5  \label{eqn:stitchedSignal}
\end{cases}
\end{eqnarray}.

This `stitched' signal is then integrated over the band as shown producing the figure of merit (FOM),
\begin{eqnarray}
\rm{FOM} & = & \int_{0.001}^{1.5} |\tilde{\dot{x}}_m| ^2  df \label{eqn:FOM}
\end{eqnarray}.

\begin{figure}[!ht]
	\centering
	\includegraphics[width=0.9\linewidth]{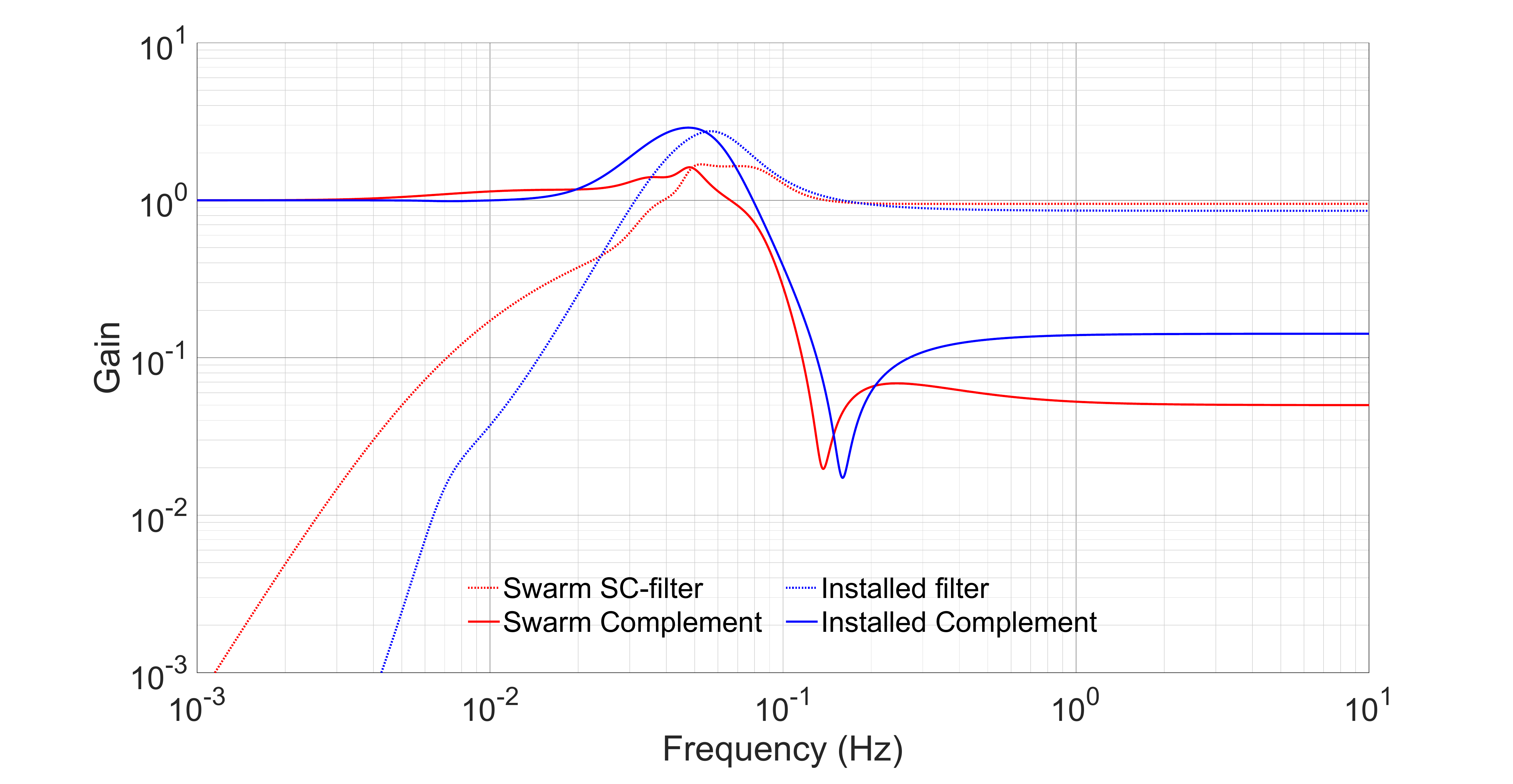}
	\caption{Figure showing a comparison between the swarmed sensor correction filter and current sensor correction filter shown in red and blue respectively. The filter itself is shown by the dashed line and its complement is shown by a solid line. Taken with permission from \cite{Cooper19PhD}.}
	\label{fig:csSCFilters}
\end{figure}
\begin{figure}[!ht]
	\centering
	\includegraphics[width=0.9\linewidth]{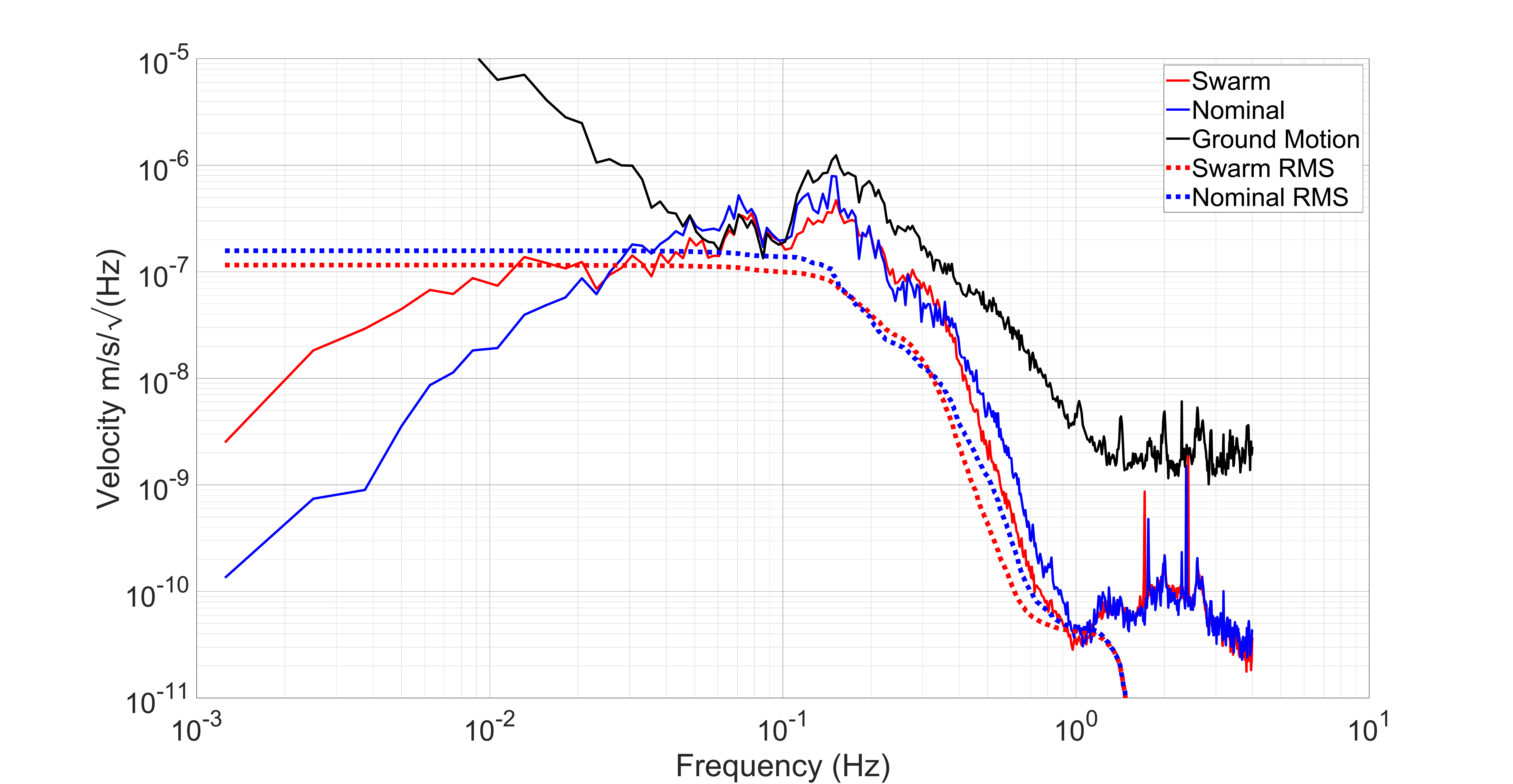}
	\caption{The cost given by Equation~\ref{eqn:FOM} for two different sensor correction filters while running on HAM2 in the X degree of freedom. The RMS velocity of each of the filters is given by the dashed traces.}
	\label{fig:HAM2XPerf}
\end{figure}

A comparison of the swarmed sensor correction filter and the current filter is shown in Figure~\ref{fig:csSCFilters}. The swarmed filter has significantly less gain peaking around 50\,mHz resulting in more tilt injection below 10\,mHz. 
Figure~\ref{fig:HAM2XPerf} shows a comparison between the two sensor correction filters when evaluated on HAM2 in the X degree of freedom.
To compensate for the differences in input ground motion when the filters were installed, the figure of merit term, is weighted by the ratio of the signal from the ground seismometers during the two times. The RMS of each of the traces are calculated from 1.5\,Hz to prevent the large spikes at 1.7 and 2.2\,Hz due to the suspension resonances from saturating the RMS. Overall, the swarmed sensor correction filter results in 27\% lower RMS velocity of the ISI compared to the current filter when the input ground motion has been normalised. While the swarmed filter causes a factor 10 more tilt injection at 1\,mHz this does little to affect the overall RMS of the platform. The swarmed filter makes most of its improvement between 0.4 and 0.9\,Hz, of particular interest is the performance difference at 0.45\,Hz, one of the suspension resonances, where the swarmed filter results in 33\% less motion than the current filter.



\begin{figure}[!ht]
	\centering
	\includegraphics[width=0.9\linewidth]{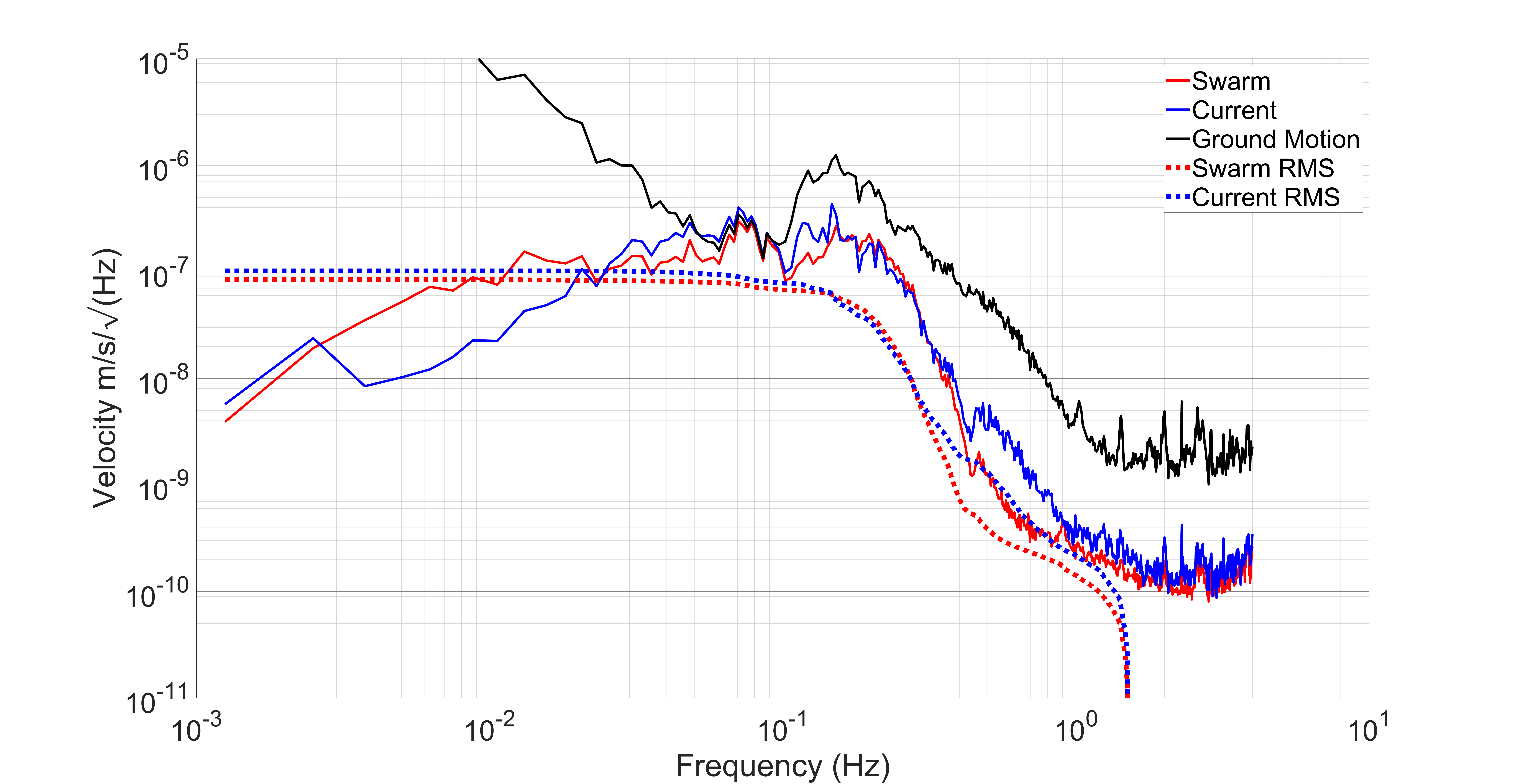}
	\caption{Figure shows the cost given by Equation~\ref{eqn:FOM} for two different sensor correction filters while running on ITMX in the X degree of freedom. The RMS of each of the filters is given by the dashed traces.}
	\label{fig:ITMXXperf}
\end{figure}

Figure~\ref{fig:ITMXXperf} shows a comparison between the swarmed and current sensor correction filter on ITMX in the X degree of freedom. The swarmed filter results in a more modest 18\% reduction in RMS velocity of the ISI. Again, the swarmed filter causes more tilt injection between 30 and 200\,mHz, though this has little effect on the overall velocity RMS. The biggest difference between the two filters comes between 0.4 and 0.9\,Hz. At 0.45\,Hz the swarmed filter results in 68\% less motion than the current sensor correction filter.  

\section{Conclusions}
Many strategies are being developed to decrease lock loss due to ground motions at the LIGO facilities. 
Improving the control filter design presents an opportunity to make significant gains with simple and affordable changes.
A robust working tool to design sensor correction filters has been developed, and filters created by this method have been deployed on-site.
This work focused on one specific filter although its design principles can be adapted for other filters. The blend filters present in all the isolation systems are excellent candidates for testing.
Designing control filters in this manner presented some challenges that had to be overcome, but techniques were developed, trialled, and tested throughout have solved many of the issues and may prove relevant to other applications. 
With this tool, better seismic isolation can be achieved leading to more observing time, more source detections and, ultimately, a greater understanding of some of the most fascinating phenomena in the universe.
\section*{Acknowledgments}
We would like to thank Brett Shapiro and Brian Lantz for their many useful suggestions and comments. Furthermore, discussions with Will Farr were an essential step for the building of filters.

\bibliographystyle{iopart-num}
\providecommand{\newblock}{}

\end{document}